\documentclass[aps,pra,preprint,showpacs,groupedaddress,superscriptaddress,longbibliography]{revtex4-1}

\usepackage[english]{babel}
\usepackage{float}
\usepackage{graphicx}
\usepackage[caption=false]{subfig}
\usepackage{amsmath}
\usepackage{amsfonts}
\usepackage{amssymb}
\usepackage{mathtools}

\begin{document}

\title{Quantum Superexponential Oscillator}

 \author{Peter Schmelcher}
  \email{Peter.Schmelcher@physnet.uni-hamburg.de}
 \affiliation{Zentrum f\"ur Optische Quantentechnologien, Universit\"at Hamburg, Luruper Chaussee 149, 22761 Hamburg, Germany}
 \affiliation{The Hamburg Centre for Ultrafast Imaging, Universit\"at Hamburg, Luruper Chaussee 149, 22761 Hamburg, Germany}

\date{\today}

\begin{abstract}
We investigate the spectral and eigenstate properties of the quantum superexponential oscillator. 
Our focus is on the quantum signatures of the recently observed transition of the energy dependent period of
the corresponding classical superexponential oscillator. We show that the ground
state exhibits a remarkable metamorphosis of decentering, asymmetrical squeezing and the development
of a tail. Analyzing the central moments up to high order a characteristic transition
from exponentially decaying moments to increasing moments is unraveled. A corresponding spectral
analysis shows that, surprisingly, to a good approximation the spectrum is equidistant. A closer
look, however, reveals a spectral scaling behaviour below the transition point which is replaced
by irregular oscillations above the transition energy. Excited bound states are analyzed up to
the continuum threshold. We discuss future perspectives and possible experimental realizations of
the superexponential oscillator.
\end{abstract}

\maketitle

\section{Introduction} \label{sec:introduction}

Oscillators represent a key ingredient for our modelling of processes in
nature. Across disciplines individual oscillators as well as coupled systems
of many oscillators provide us with a fundamental understanding of the mechanisms underlying
the structural properties and dynamics of light and matter. This ranges from 
the statistical properties of light \cite{Scully} and the synchronization phenomenon
of coupled oscillators \cite{Arenas} to the collective motion and vibrations of atoms
in a crystal \cite{Ashcroft}.  

A distinguished starting-point is a linear system of harmonic oscillators described by
an integrable Hamiltonian which is characterized by its eigenfrequencies and normal modes.
The latter provide a very good description of the small amplitude motion around the 
corresponding equilibrium configurations for e.g. molecules \cite{Wilson} and collective atomic
behaviour in a bulk \cite{Ashcroft}. Larger amplitude motions add then nonlinearities which can be responsible
for a plethora of nonlinear phenomena, such as breathers \cite{Flach1,Flach2}, traveling excitations
in the form of solitons \cite{Kevrekidis,Kartashov} or nonlinear vibrational response properties like
phononic frequency combs through nonlinear resonances \cite{Cao}.
Exploring coupled nonlinear oscillators allows to address important questions such as the
thermalization in many-body systems which can be strongly influenced by the existence of nonlinear
excitations, approximate or exact constants of motion and the resulting near integrability \cite{FPU}.

External driving of oscillators adds significantly to the pathway from integrable to
chaotic behaviour possessing
major implications on several relevant topics e.g. the classical energy diffusion \cite{Casati}
and the corresponding quantum localization \cite{Izrailev,Stoeckmann}.
Driving offers a high degree of control on the underlying phase space thereby
engineering not only its ergodic chaotic component but in particular the appearance of 
regular islands therein. Resonances via parametric driving \cite{Fossen} lead, among many others, to
pattern formation in soft matter \cite{Arbell} and to multi-mode lasing \cite{Szwaj}.

In view of the large number of phenomena and applications associated with oscillators the 
very recent finding of superexponential oscillators is of immediate interest \cite{Schmelcher1,Schmelcher2,Schmelcher3}.
Opposite to common oscillators which are based on a power law confinement $V(q) = q^{\beta}$ with a constant
power $\beta$, the potential (SEP) of the self-interacting superexponential oscillator (SSO)
is of the appearance ${\cal{V}} (q) = |q|^q$ \cite{Schmelcher2}. 
The SEP is therefore highly nonlinear with the dynamical degree
of freedom appearing both in the base as well as the exponent. It possesses a pronounced asymmetric potential well
with a stable equilibrium and the underlying phase space features a region of agglomerating
curves. It has been shown \cite{Schmelcher2}, among others, that the
potential well of the SSO exhibits 
two regimes of different dynamical behaviour: for low energies its period $T(E)$ scales (approximately) linearly
with the energy $E$ whereas above a transition point characterized by a singular behaviour of the derivatives of
the SEP, the SSO shows a highly nonlinear functional dependence of $T(E)$. These features are not known from common
oscillators and are unique to the SSO.

It is therefore a natural question to ask what the quantum signatures of the above-mentioned properties
of the SSO are. Does the crossover observed in the classical SSO have a corresponding counterpart
in the quantum SSO ? What are its spectral properties and what is the structure of the ground and
excited states with varying parameters ? These are the central questions we will answer in this
work. While the ground state energy passes the transition point of the SEP with varying amplitude
it experiences a decentering, compression and tail development.
We characterize the crossover behaviour of the ground state by identifying corresponding features in the
central moments of its probability distribution. A spectral analysis reveals the amazing fact
that in lowest order approximation there is an equidistant spacing of the energy levels. However, 
a closer look shows that a characteristic spectral scaling behaviour exists below the transition point.
Above the transition point energy this scaling is lost and irregular fluctuations take over. The higher excited
eigenstates are also analyzed in some detail.

We proceed as follows. Section \ref{sec:setup} contains a description of the SSO main characteristics.
Section \ref{sec:gs} is dedicated to the investigation of the ground state of the SSO with varying 
amplitude of the SSO potential. This includes the ground state crossover via the transition point 
of the SSO potential. A main tool for the characterization of the ground state is an analysis of 
the probability distribution via its central moments. In section \ref{sec:spectrum} we explore
the spectra focusing on the case of many bound states. An investigation of the energetical spacing
and the related scaled spacing with increasing degree of excitation is performed. An eigenstate analysis
for excited states is presented in section \ref{sec:esa}. Finally we give our conclusions in 
section \ref{sec:conclusions}. Appendix \ref{sec:appendix} contains a brief discussion of the
spectral properties of arbitrary power law potentials for reasons of comparison to the SSO.

\section{The self-interacting superexponential oscillator} \label{sec:setup}

Commonly used oscillators possess a power law potential, such as the harmonic, quartic or
linear potential based oscillators. Their spectral behaviour is well-understood and has been exploited and
applied in a variety of physical situations (see refs. \cite{Bender,Znojil,Quigg,Liverts} and refs. therein and
the appendix \ref{sec:appendix} of the present work).
Here we explore the quantum physics of a novel type of oscillator, the self-interacting superexponential
oscillator, whose classical properties and dynamics have been investigated in a recent work \cite{Schmelcher2}.

\begin{figure}
\hspace*{-4cm} \parbox{9cm}{\includegraphics[width=12.4cm,height=6.5cm]{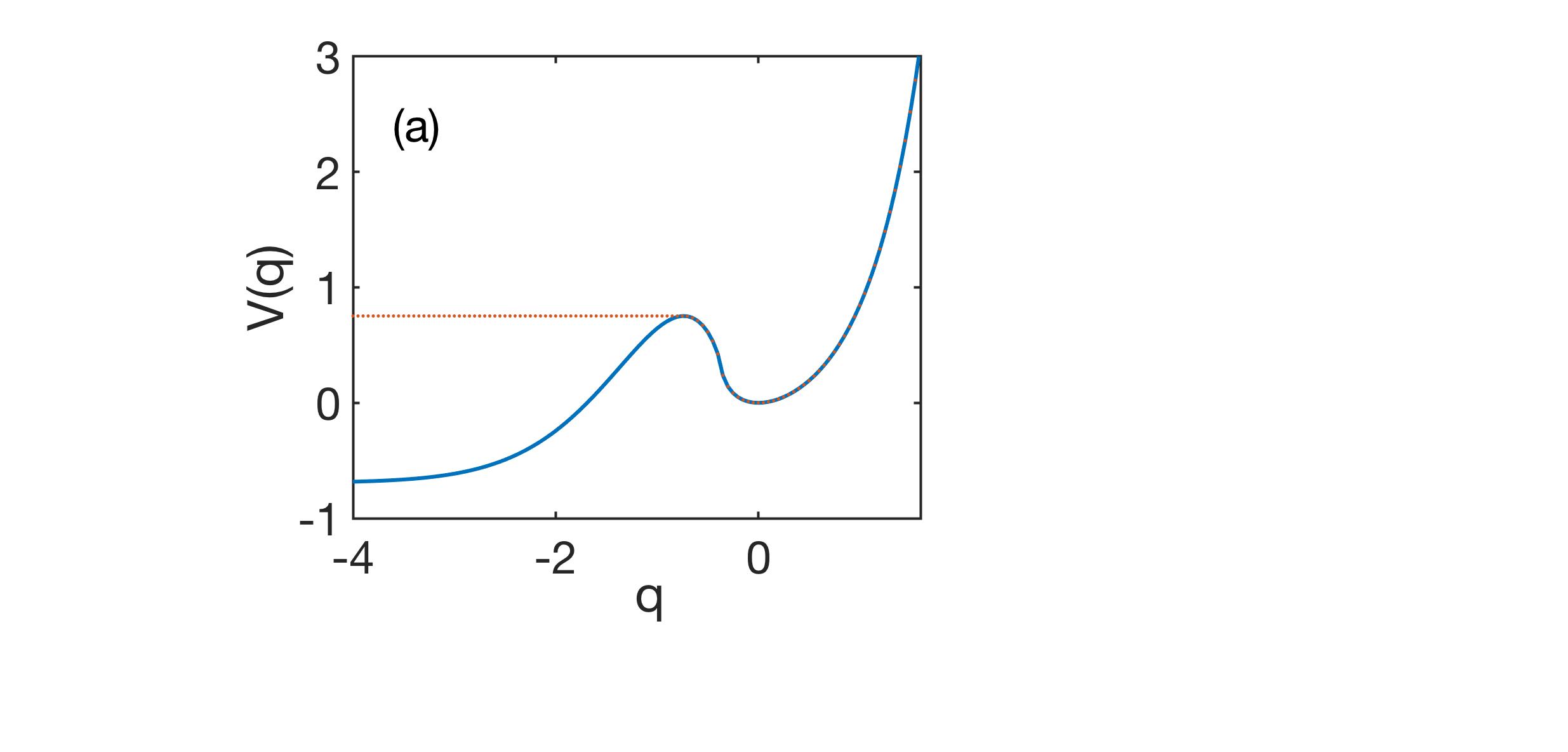}}
\hspace*{-1.4cm} \parbox{7cm}{\vspace*{0.2cm} \includegraphics[width=11.5cm,height=6.8cm]{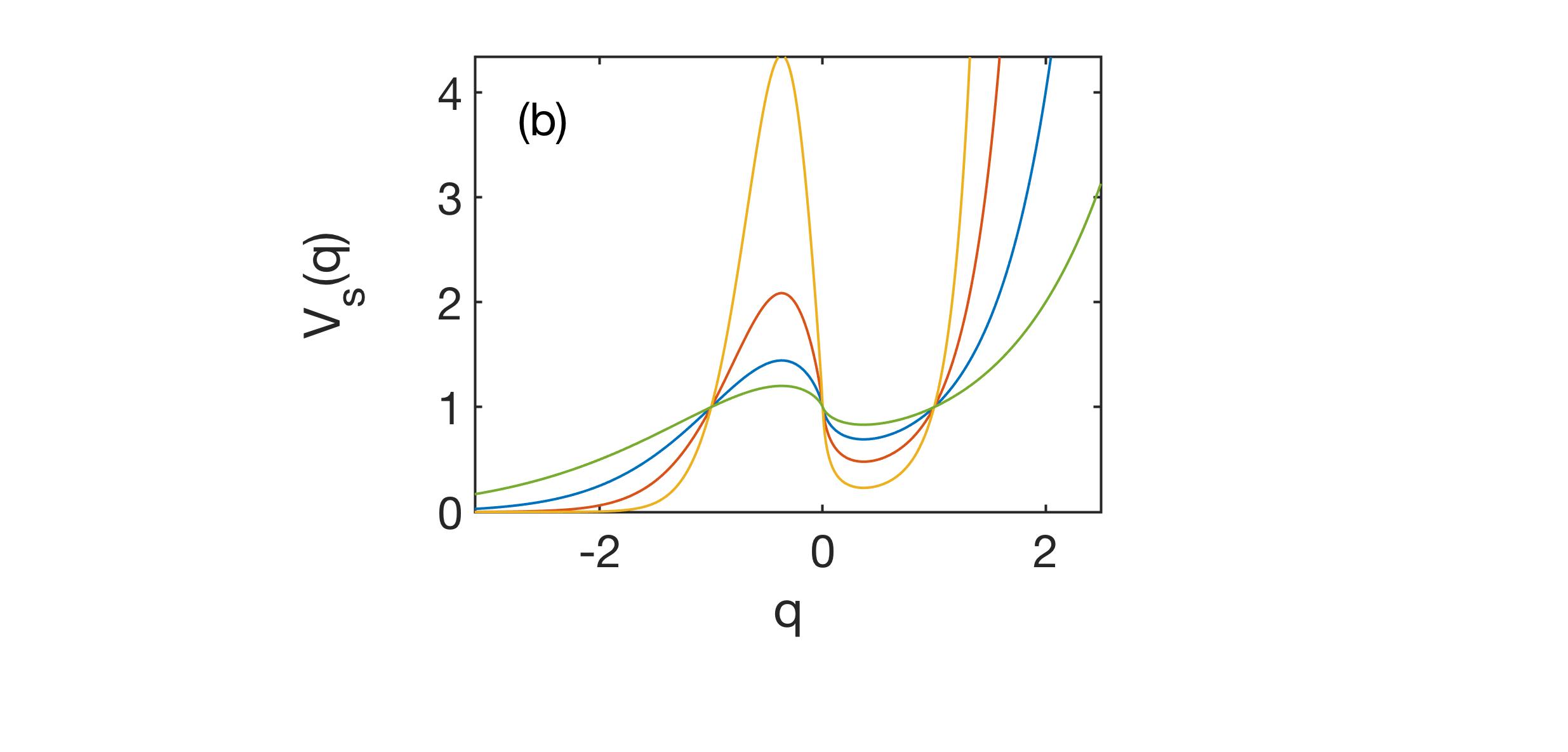}} \vspace*{-1cm}
\caption{(a) The spatially and energetically shifted potential of the SSO, see eq.(\ref{shiftedsep}).
The dotted horizontal line indicates the constant behaviour of ${\cal{V}}_m$ for $q \leq - \frac{2}{e}$.
(b) Skewed SSO potential ${\cal{V}}_s(q)= \alpha |q|^{\delta q}$ for different values
of the parameter $\delta$. Here $\alpha = 1$ and $\delta = 0.5, 1, 2, 4$
correspond to the curves with increasing energy at e.g. $q=2$.}
\label{Fig1}
\end{figure}

The Hamiltonian of the SSO reads as follows

\begin{equation}
{\cal{H}} = \frac{p^2}{2m} + \alpha |q|^q , \hspace*{0.5cm} \alpha >0, m>0
\label{Hamiltonian}
\end{equation}

with the SEP ${\cal{V}} = \alpha |q|^q$ where $\alpha$ is a corresponding amplitude
and $m$ the mass of the oscillator. The resulting stationary Schr\"odinger equation reads

\begin{equation}
\left( - \frac{\hbar^2}{2m} \frac{d^2}{dq^2} + \alpha |q|^q \right) \Psi_n = E_n \Psi_n
\label{SEQ}
\end{equation}

whose eigenstates and spectral properties are of central interest in the present work.
Opposite to the commonly employed power law oscillators with a constant power \cite{Bender,Znojil,Quigg,Liverts} 
the SEP depends on the dynamical degree of freedom $q$ not only in case of the base but also
in the exponent. This very unusual superexponential and self-interacting appearance leads to a strong
spatial variation of its highly nonlinear character. As a result the SSO exhibits a number of peculiar 
properties, as explored in ref.\cite{Schmelcher2} for the classical case. 
Let us briefly summarize some of the most important features of the SSO. As presented in Figure \ref{Fig1}(a) 
it shows a repulsive wall of superexponential nature for $q \rightarrow +\infty$, separated by a 
potential well with a corresponding minimum and maximum from a decaying asymptotically constant
behaviour for $q \rightarrow - \infty$. The potential well of the SSO is highly nonlinear and asymmetric and
possesses in particular no reflection symmetry around its minimum. The corresponding minimum and maximum are placed
at $q_{min}= \left( \frac{1}{e} \right)$ with ${\cal{V}}(q_{min})=e^{-\frac{1}{e}}$ and for
$q_{max}= \left(- \frac{1}{e} \right)$ we have ${\cal{V}}(q_{max})=e^{\frac{1}{e}}$.
Symmetrically placed between the minimum and the maximum there is a so-called transition point 
which has been analyzed in detail in ref.\cite{Schmelcher2}: all derivatives of the SEP are singular
at this point implying logarithmic and power law singularities of increasing power with an increasing
order of the derivative. A Taylor expansion around the minimum reveals that the $N$-th derivative
dominant term scales with $e^{N-1}$ showing again the major difference of the SSO compared to any common
power law oscillator, such as the harmonic oscillator. One of the main features of the
dynamics in the potential well of the SSO is the peculiar dependence of its
period on the energy. For energies below the transition point an approximately linearly
decreasing behaviour is encountered whereas for energies above the transition point a nonlinear
increasing behaviour occurs. This is again very different from standard power law oscillators for
which $T(E) \propto E^{-1/2+1/2n}$ with $V \propto q^{2n}, n \in \mathbb{N}$. It is therefore
a natural and intriguing question to ask what the quantum behaviour and properties of the SSO are.

Let us provide some basic facts for our exploration of the spectrum and eigenstates of the quantum SSO. 
First of all we note that the Hamiltonian (\ref{Hamiltonian}) possesses a scaling property concerning its
dependence on the mass $m$ and amplitude $\alpha$ which leads to the relation 
$E(\beta,1)=E(1,\beta) \cdot \frac{1}{\beta}$ with $\beta= \alpha m$ where the two arguments of the
energy stand for the inverse prefactors of the kinetic energy and the potential respectively.
In the following numerical investigations we can therefore focus on varying a single parameter only
which allows us to tune the energy level density in the potential well of the SEP.

For reasons of simplification we shift the SEP to have the minimum at $q=0$ and the
corresponding energy value being zero. This yields

\begin{equation}
{\cal{V}} (q) = \alpha \left( |q+e^{-1}|^{(q+e^{-1})} - e^{-\frac{1}{e}} \right)
\label{shiftedsep}
\end{equation}

We have then $q_{min}=0$ and $q_{max}=-\frac{2}{e}$ and 
${\cal{V}}_{max}={\cal{V}} (q_{max}) = \alpha e^{\frac{1}{e}} \left( 1 - e^{-\frac{2}{e}} \right)$.
The transition point with the vertical derivative is then located at
$q_v=-\frac{1}{e}$ with the value ${\cal{V}}_{v}={\cal{V}} (q_{v}) = \alpha \left( 1 - e^{-\frac{1}{e}} \right)$.
As a next step we define the modified version ${\cal{V}}_m$ of the SSO potential such that it possesses exact
bound states and no outward tunneling to the continuum for states in the confining potential well occurs
(see Fig.\ref{Fig1}(a) for $q<q_{max}$).

\[
    {\cal{V}}_m(q)=  
\begin{cases}
    \alpha \left( |q+e^{-1}|^{(q+e^{-1})} - e^{-\frac{1}{e}} \right), & \text{if } q\geq -\frac{2}{e}\\
    \alpha e^{\frac{1}{e}} \left( 1 - e^{-\frac{2}{e}} \right),       & \text{otherwise}
\end{cases}
\]

The above modification has only a minute impact and leads only to very minor changes 
to the bound states in the potential well of the SEP. Exceptions are those states which are energetically
very close to the maximum (see below).

Finally we remark that variations of the SEP can be obtained by considering a skewed SEP which 
reads ${\cal{V}}_s(q) = \alpha |q|^{\delta q}$. This includes an additional parameter 
$\delta$ which allows to scale the behaviour in the exponent. Fig. \ref{Fig1}(b) shows the
skewed SEP for different values of $\delta$. Overall, this can lower or increase the 
well depth, while the asymmetry of the overall well is maintained, but it is stretched 
in a spatially dependent manner. In the following we focus on $\delta = 1$.

Our numerical approach to solve the stationary Schr\"odinger equation (\ref{SEQ}) is an eighth order
finite difference discretization scheme of space \cite{Groenenboom}. Determining the eigenvalues
and eigenvectors then corresponds to the diagonalization of the resulting Hamiltonian band matrix.
We use $\hbar=m=1$ herefore.

\section{Ground state of the SSO} \label{sec:gs}

\begin{figure}
\parbox{9cm}{\includegraphics[width=11cm,height=6cm]{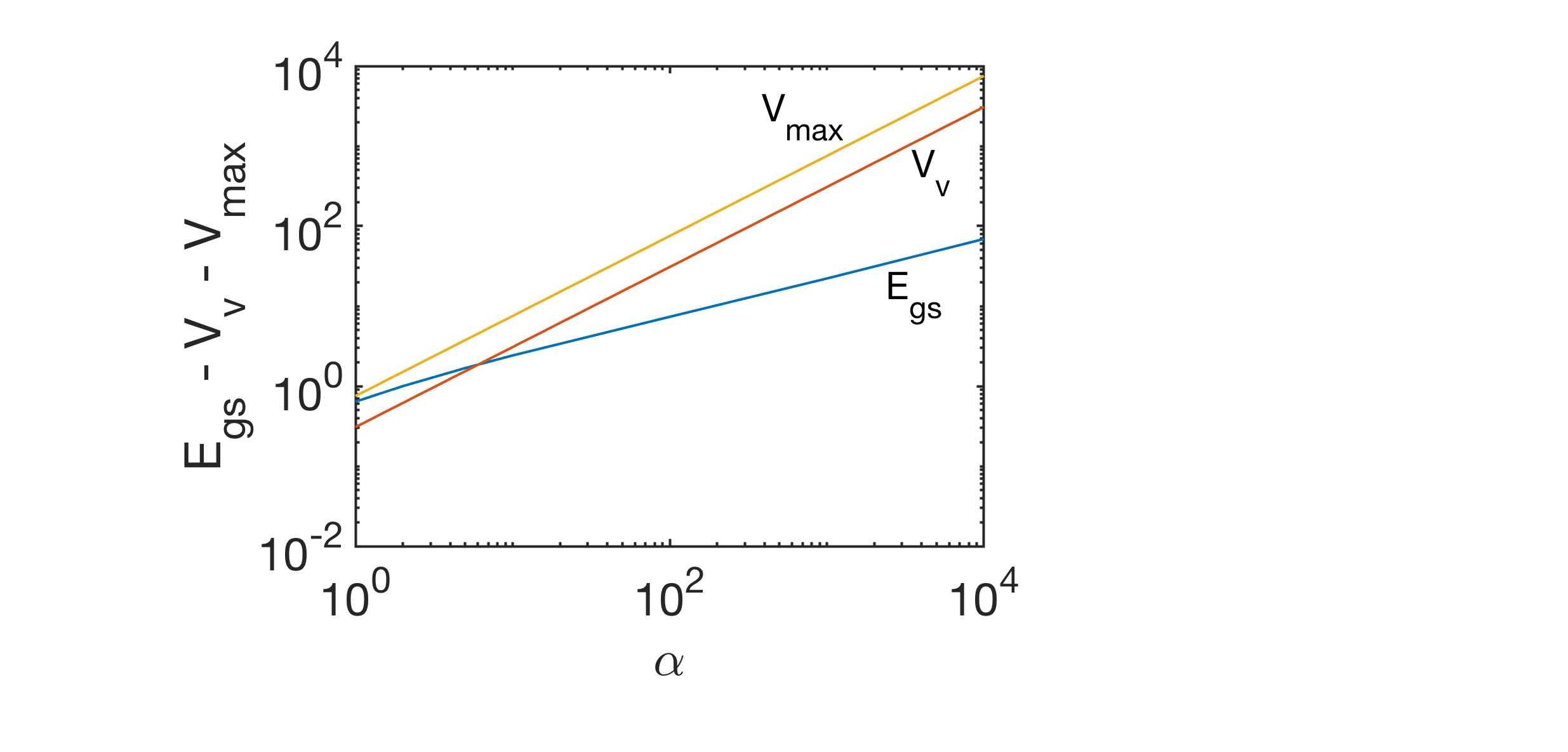}}
\caption{The energy of the ground state $E_{gs}$ of the SSO together with the energy of the maximum ${\cal{V}}_{max}$
and transition point ${\cal{V}}_{v}$ as a function of $\alpha$ on a double logarithmic scale.}
\label{Fig2}
\end{figure}

In this section the focus is on the ground state of the SSO with varying parameter $\alpha$,
the amplitude of the SSO potential. According to the above-given scaling relation of the
energy changing $\alpha$ is equivalent to changing the mass of the SSO and varies the
density of states and consequently the number of bound states in the potential well of the SSO. 
Here, we will vary $\alpha$ such that the energy of the ground state moves from being very close
to the minimum of the well to being close to the energy of the maximum that defines the
escape to the continuum. Figure \ref{Fig2} shows this 'motion' together with the energies
of the maximum and of the transition point of the SEP.

For large values of $\alpha$ the ground state energy is below ${\cal{V}}_v (\alpha)$ (the latter is the transition point
of singular derivatives) and for small values of $\alpha$ it is above ${\cal{V}}_v (\alpha)$.
These regimes are clearly discernible in Fig.\ref{Fig2} together with the crossover that 
takes place at $\alpha \approx 6$. The energies ${\cal{V}}_{v}(\alpha),{\cal{V}}_{max}(\alpha)$ show the same
dependence on $\alpha$ for large values of $\alpha$ (see parallel curves in Fig.\ref{Fig2}). 

\begin{figure}
\parbox{9cm}{\includegraphics[width=11cm,height=6cm]{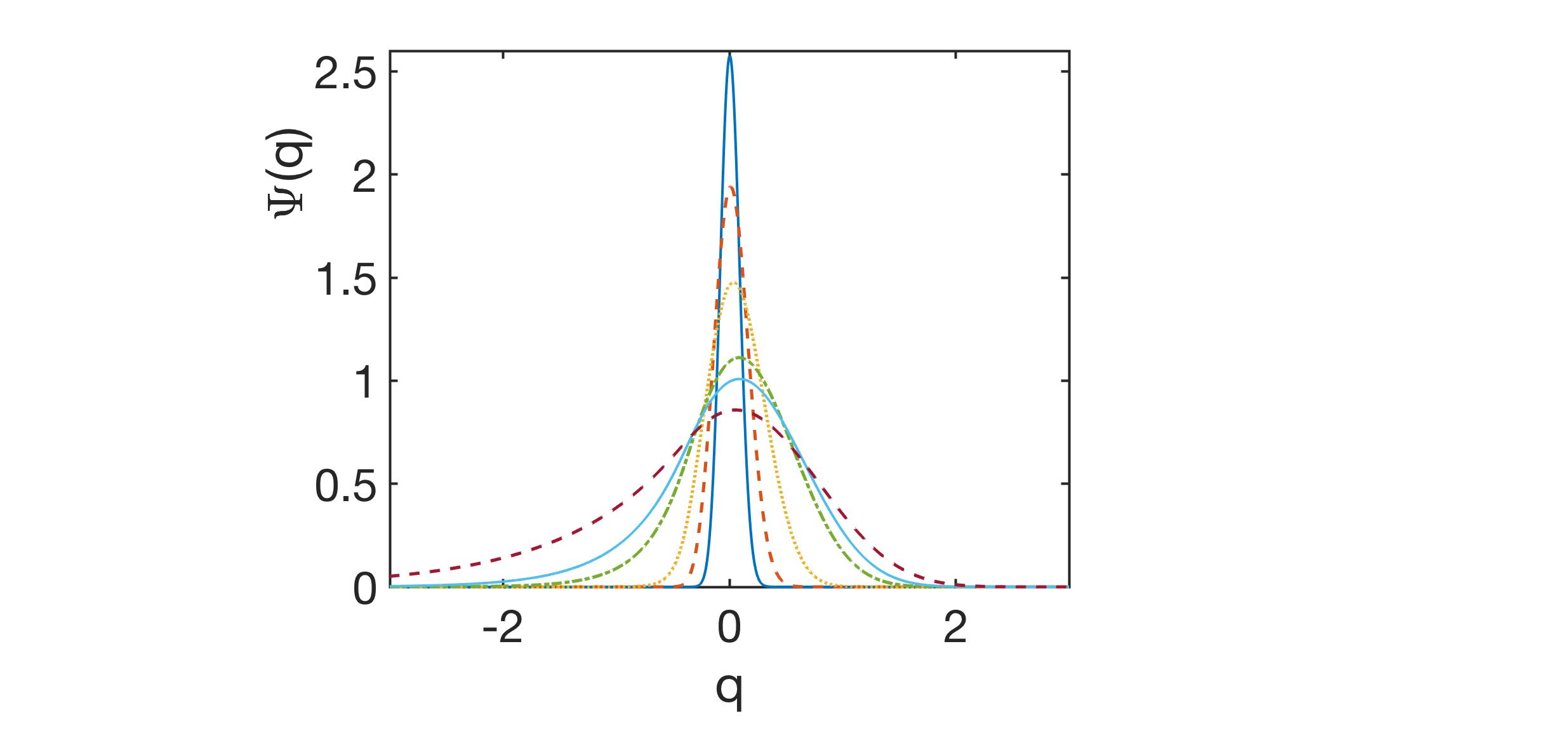}}
\caption{The wave function of the ground state $\Psi(q)$ for $\alpha = 10^4$ (solid), $10^3$ (dashed), $10^2$ 
(dotted), $10$ (dash-dotted), $5$ (solid), $2$ (dashed). A decreasing value of $\alpha$ 
implies an increasing width of $\Psi (q)$.}
\label{Fig3}
\end{figure}

Figure \ref{Fig3} shows the ground state wave function for the values $\alpha=10^4,10^3,10^2,10,5,2$
of the amplitude.
First of all we note that due to the lack of a reflection symmetry at $q=0$ the maximum of the ground 
state is not located at $q=0$ but decentered. Consequently there is no parity symmetry and the
ground (as well as excited states, see section \ref{sec:esa}) state exhibits an asymmetric profile. 
Furthermore with a decreasing value of $\alpha$ a half-sided compression due to the superexponential
wall for $q \rightarrow +\infty$ occurs accompanied by the formation of a tail for negative values of $q$.
The major change of the ground state character happens when it passes the transition point energy. To analyze this
in more detail we investigate the behaviour of the central moments 

\begin{equation}
m_n = \int_{-\infty}^{+\infty} \left( q - \mu \right)^n |\Psi|^2 (q) dq
\label{moments}
\end{equation}

of the probability distribution of the ground state with varying amplitude $\alpha$.
$\mu$ is the mean $<q>$ and $m_0$ is one due to normalization whereas $m_1$ is zero due
to the centrality of the moment. Generally speaking the mean $\mu$ and the variance $m_2$
provide information on the location (centering) and the variability (spread or dispersion)
of the probability distribution. The third central moment $m_3$ is linked to the skewness 
which is a measure for the left right asymmetry or parity violation. Reflection symmetry
implies zero skewness. Tails left or right lead to negative or positive skewness tentatively.
The fourth central moment $m_4$ is related to the kurtosis which is a measure for the
flatness or peakedness of a distribution.

\begin{figure}
\hspace*{-4cm} \parbox{9cm}{\includegraphics[width=11cm,height=6cm]{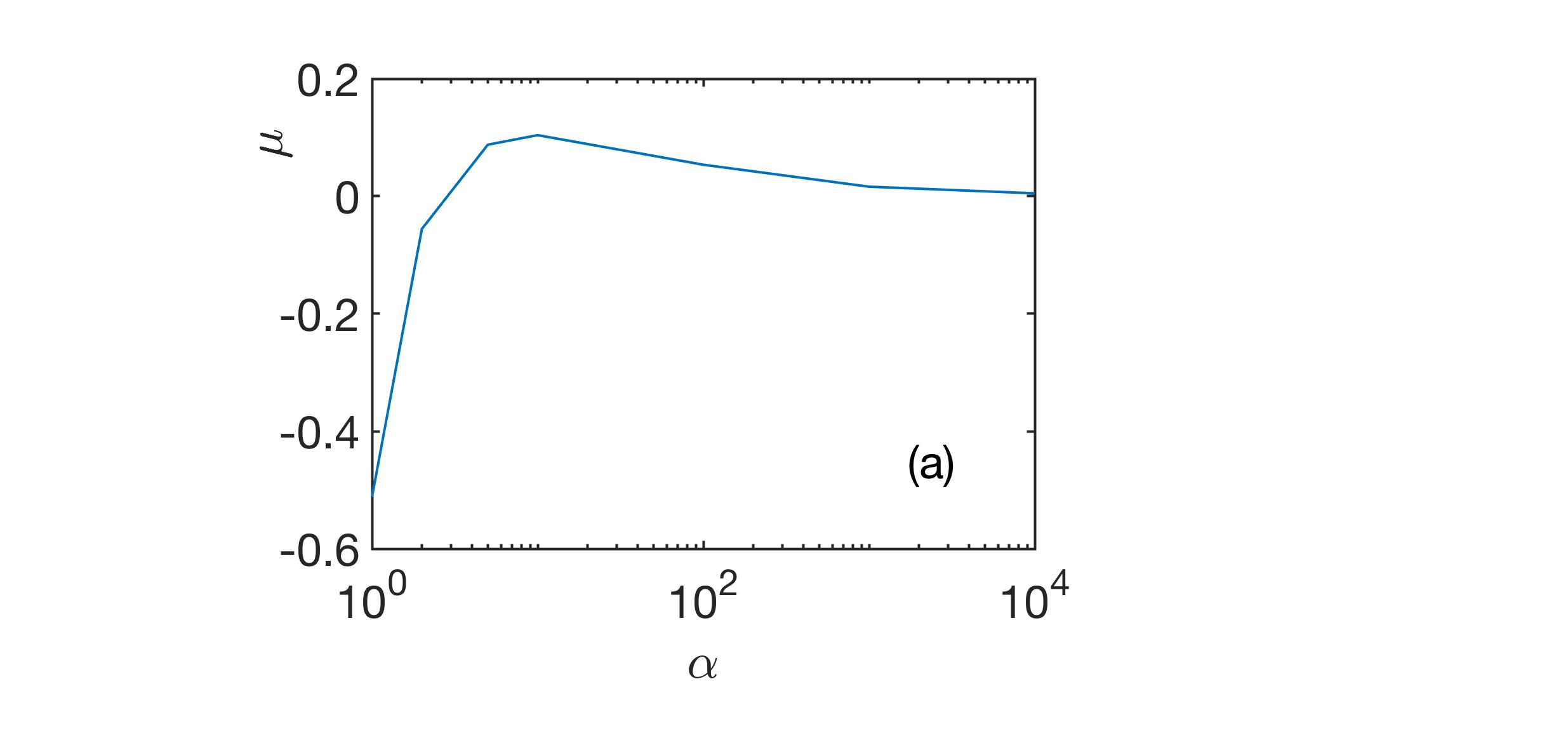}}
\hspace*{-1.4cm} 
\parbox{7cm}{\includegraphics[width=12.5cm,height=6.0cm]{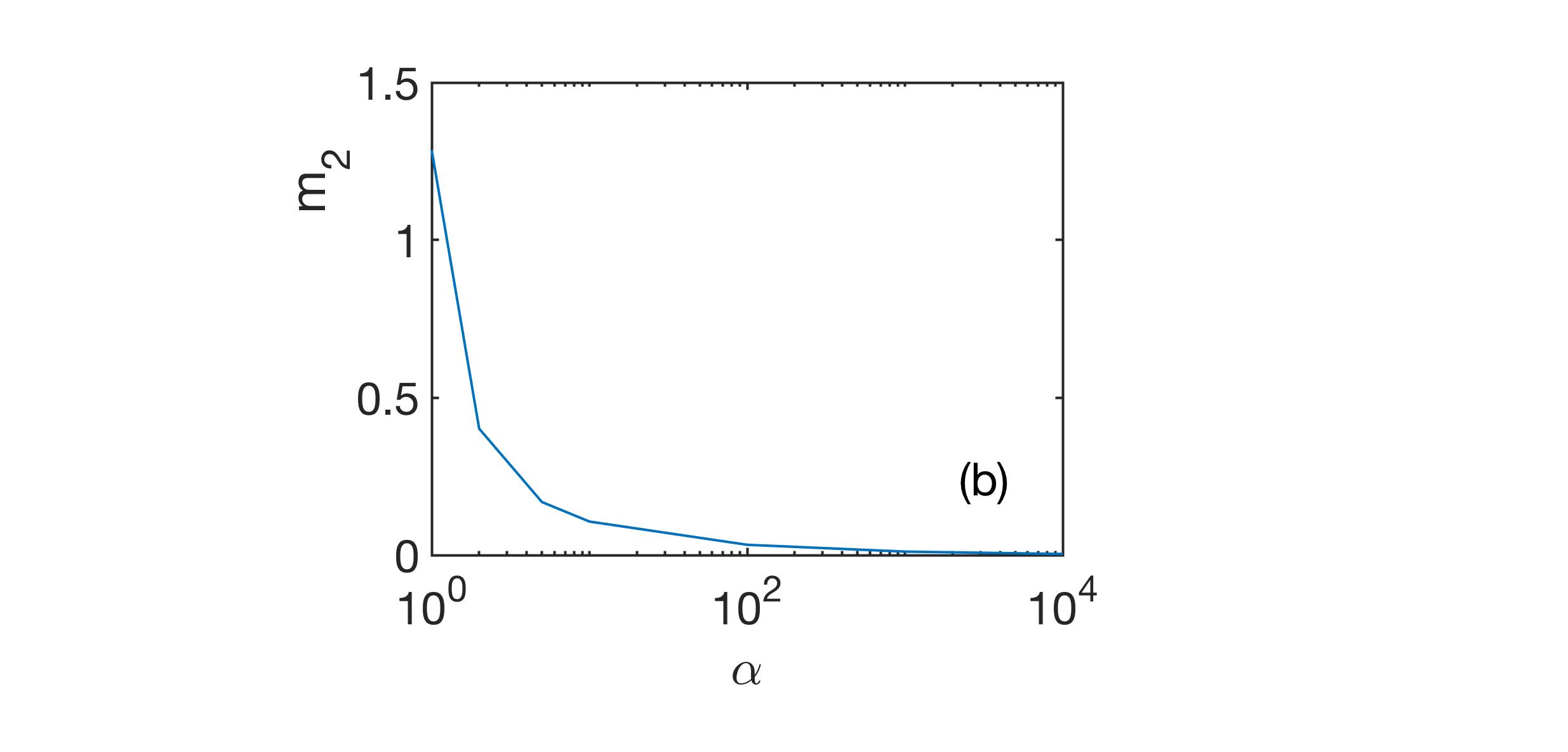}}
\vspace*{-0.5cm}
\caption{(a) The mean value $\mu = <q>$ for the ground state of the SSO with varying amplitude $\alpha = 1-10^4$.
(b) The variance $m_2=<(q-\mu)^2>$ for the ground state with varying $\alpha$.}
\label{Fig4}
\end{figure}

Fig.\ref{Fig4}(a,b) show the mean value $\mu = <q>$ and the variance $m_2$ for the ground state of
the SSO with varying amplitude $\alpha$ respectively. For $\alpha = 10^4$ the mean is positive with a value of
approximately $0.005$ whereas the variance is approximately $0.06$. With decreasing value
for $\alpha= 10^3, 10^2, 10$ the mean becomes approximately $0.016,0.05,0.1$, i.e. it increases
substantially and the variance $m_2$ takes on the values $0.011, 0.033, 0.11$ which means
it also increases significantly. Following up on a maximum ($\alpha \approx 10$) and for further decreasing $\alpha$
the mean expectation value now decreases to $0.087$ for $\alpha = 5$ while the corresponding 
variance still increases to $0.17$ approximately.
For $\alpha = 2, 1$ a decentering to negative mean values $-0.05, -0.5$ takes place and the 
variance rapidly increases further to $0.4, 1.3$.
To summarize, the above shows that for large values of $\alpha$ a decentering occurs to the right
half of the well (towards the superexponential wall) which turns into a negative decentering to the left half of the
well for small values of $\alpha$. The crossover between the two behaviours occurs when the ground
state energy passes the transition point energy which happens at $\alpha \approx 6$.

\begin{figure}
\hspace*{-4cm} \parbox{9cm}{\includegraphics[width=11cm,height=6cm]{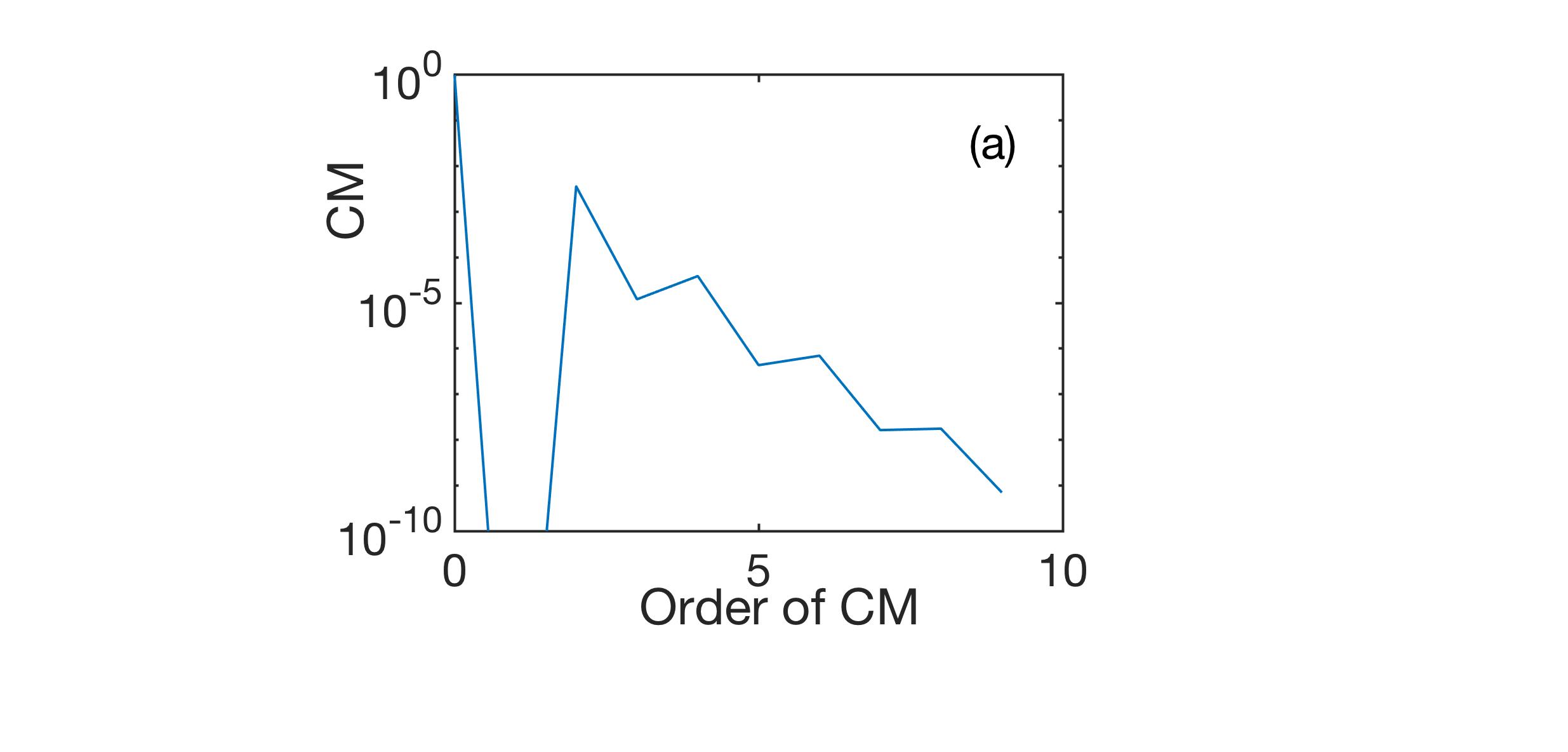}}
\hspace*{-1.6cm} 
\parbox{7cm}{\vspace*{0.2cm} \includegraphics[width=11.7cm,height=6.4cm]{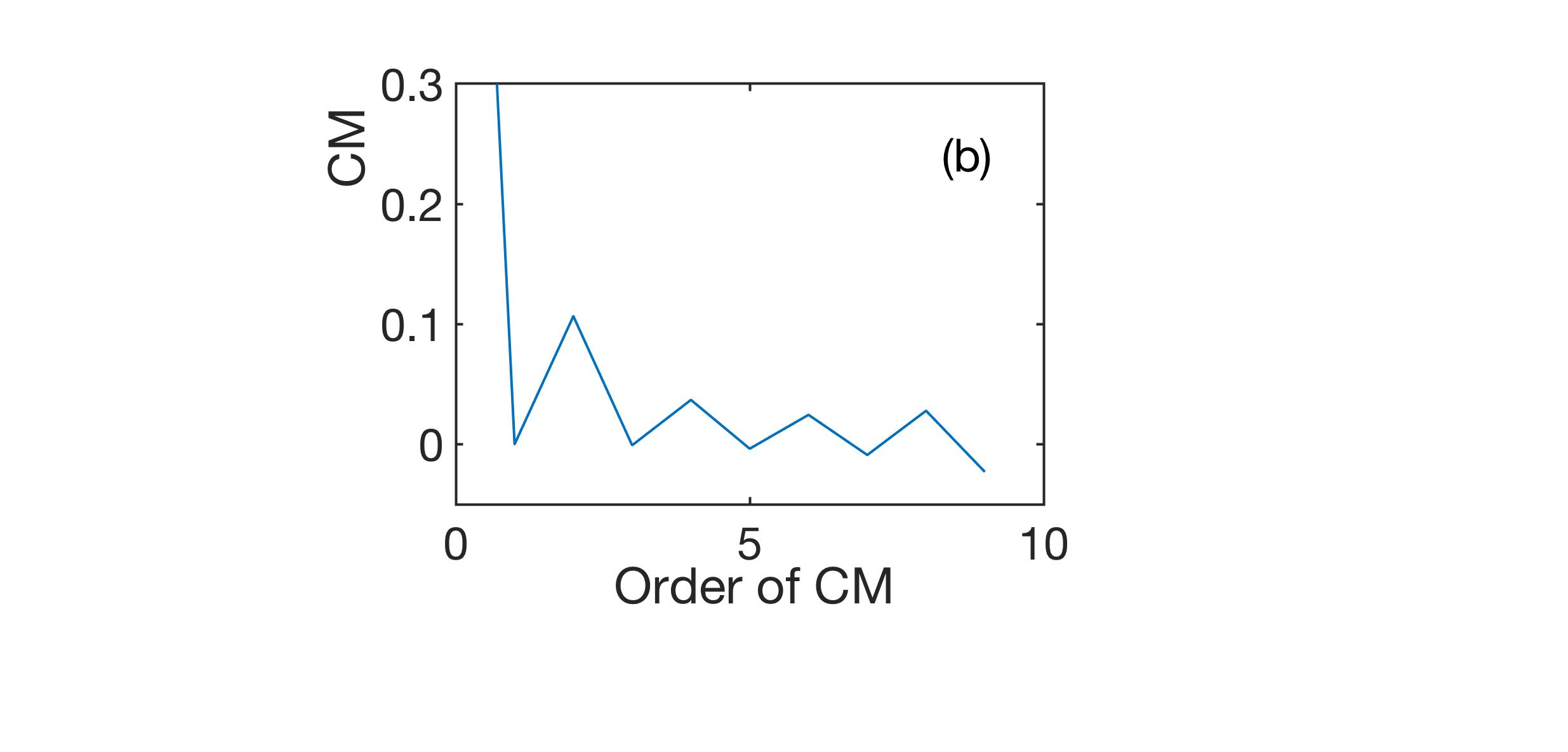}}\\
\hspace*{-3.7cm} \parbox{9cm}{\vspace*{-1.1cm} \includegraphics[width=11.6cm,height=6cm]{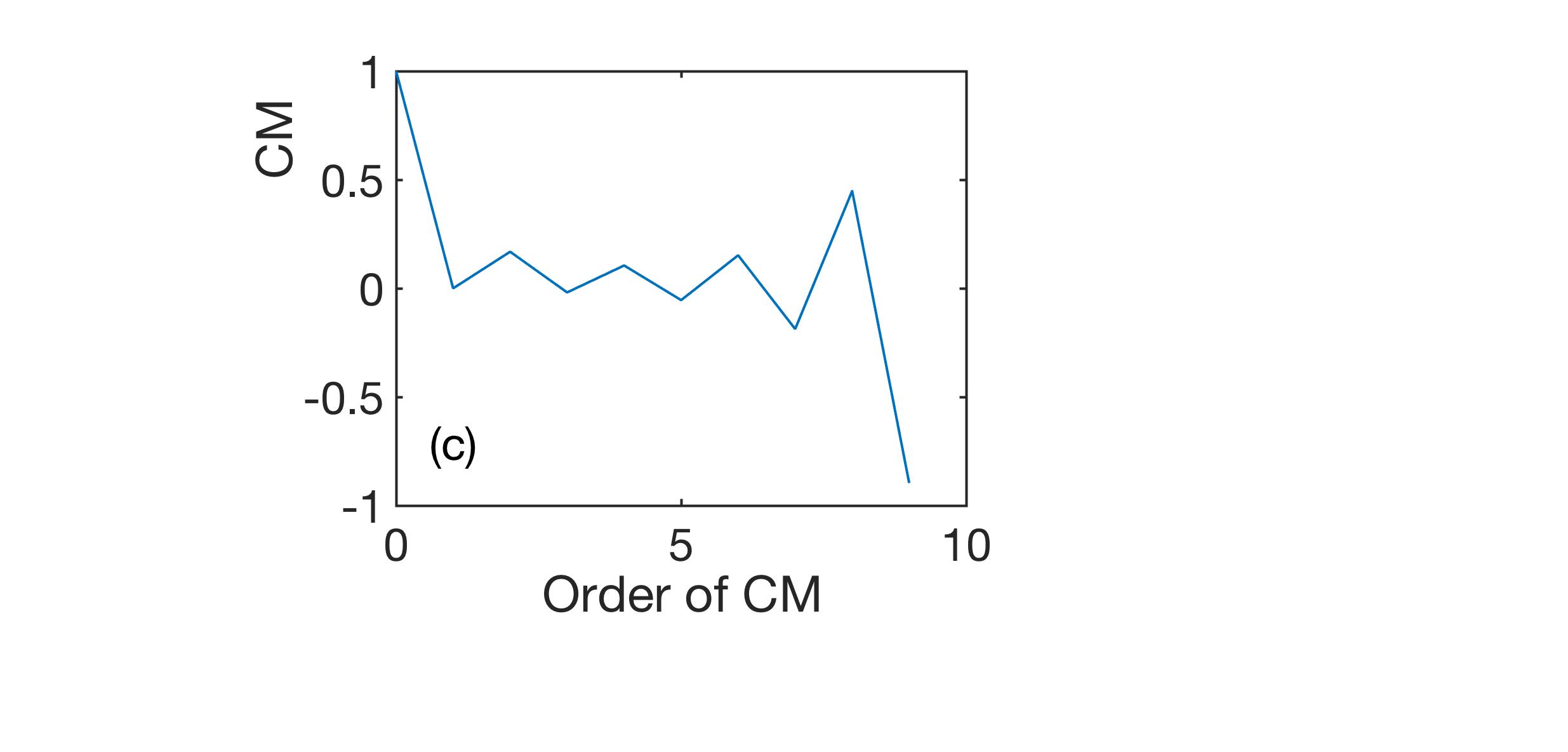}}
\hspace*{-2.1cm} 
\parbox{7cm}{\vspace*{-0.8cm} \includegraphics[width=11.7cm,height=6.0cm]{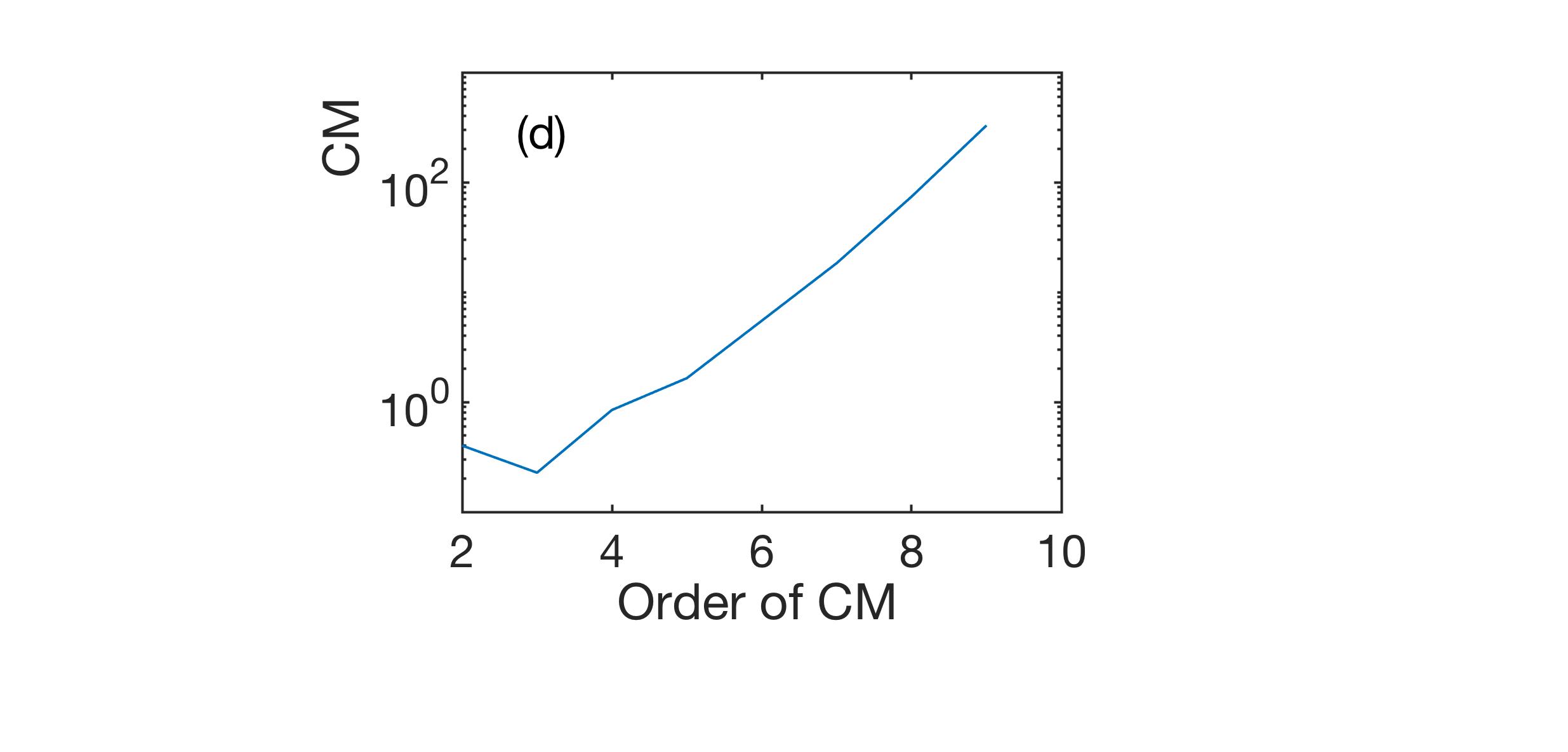}}
\caption{Central moments up to nineth order for the ground state of the SSO for (a) $\alpha = 10^4$
(b) $\alpha = 10$ (c) $\alpha = 5$ and (d) $\alpha = 2$ showing the magnitude of the central moments
on a logarithmic scale.}
\label{Fig5}
\end{figure}

Figure \ref{Fig5} shows the higher central moments up to nineth order for the values $\alpha = 10^4, 10, 5, 2$.
For $\alpha = 10^4$ (see Fig.\ref{Fig5}(a)) we observe strongly decaying central moments with increasing 
order. On top of this exponential decay there is an alternating even odd oscillation where the odd central moments
are always somewhat smaller compared to the even ones. Therefore, the skewing is smaller as compared to the
overall symmetric 'shaping'. The same behaviour can be observed for $\alpha = 10^3, 10^2$ (not shown here)
but the values of the moments overall increases significantly.

Figure \ref{Fig5}(b) addresses the case $\alpha = 10$. The decay of the central moments is now much weaker,
as compared to the above cases (note the linear scale) and the alternating even odd sequence
leads to a sequence of positive and negative values of the even and odd central moments.
The latter characterizes the emerging tails of the wave function (see Fig.\ref{Fig3}).
For $\alpha = 5$ in Fig.\ref{Fig5}(c), remarkably,
the central moments grow with increasing order and also alternate between positive
and negative values. This behaviour is even more pronounced for the case $\alpha =2$ in Figure \ref{Fig5}(d)
which shows the magnitude of the central moments on a logarithmic scale demonstrating their exponential
increase with increasing order. Very large values for higher central moments occur therefore.
The above analysis clearly demonstrates that the central moments carry not only the information
about the decentering and characteristic asymmetry of the ground state but in particular they
exhibit the fingerprints of the crossover from energies below the transition point and above it. 


\section{Spectral analysis of the SSO} \label{sec:spectrum}

We will now focus on the spectral properties of the SSO, i.e. the structure and properties
of the eigenvalue spectrum. As a representative example we choose the amplitude $\alpha = 10^4$
which leads to 48 bound states in the potential well of the SSO.

\begin{figure}
\hspace*{-4cm} \parbox{9cm}{\includegraphics[width=14.4cm,height=7.5cm]{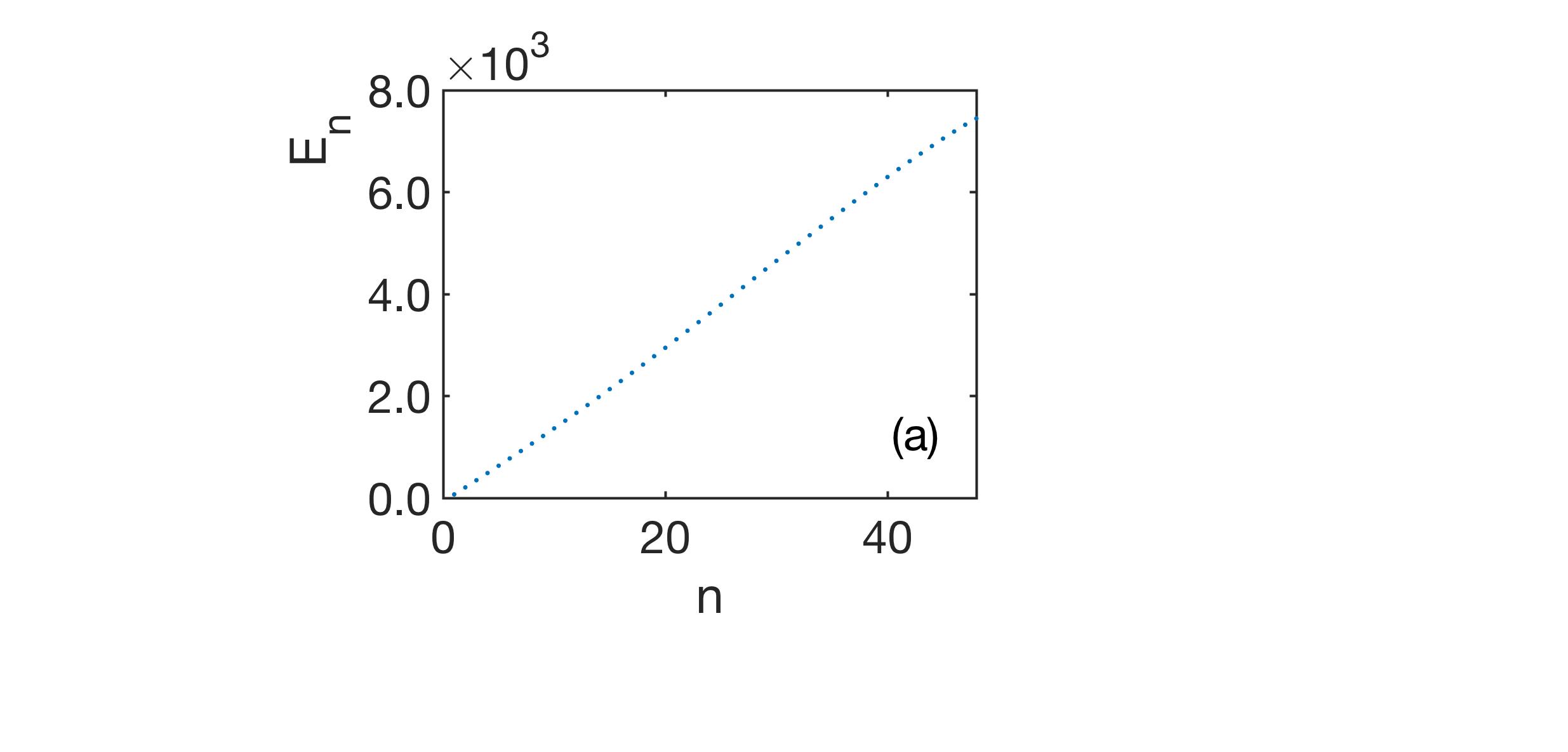}}
\parbox{7cm}{\vspace*{-0.1cm} \includegraphics[width=11.7cm,height=6.4cm]{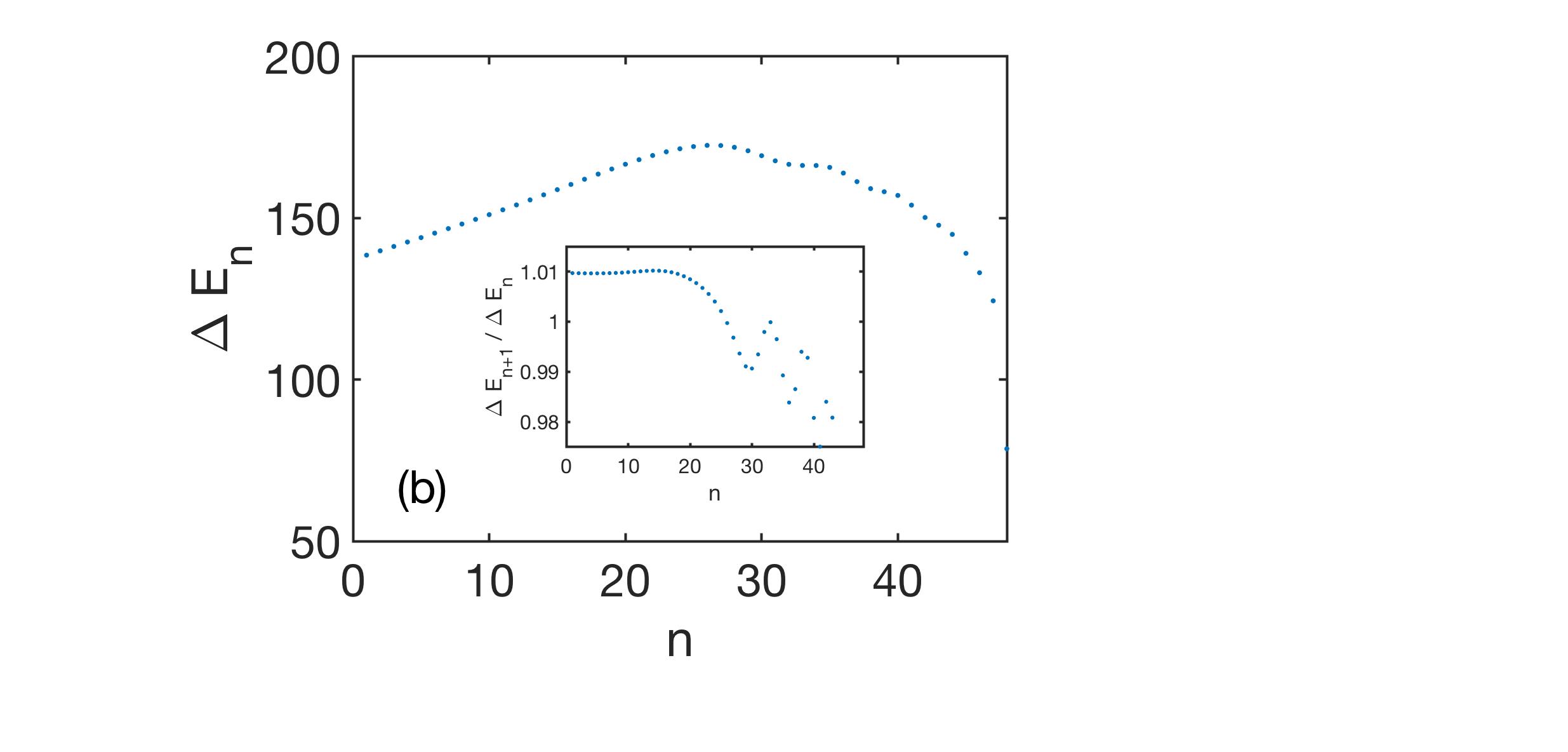}}\\
\caption{(a) The energy eigenvalue spectrum $E_n$ for $\alpha = 10^4$. (b) The spacing
$\Delta E_{n} = E_{n} - E_{n-1}$ as a function of the degree of excitation $n$. Inset: The scaled
spacing $\Delta E_{n+1} / \Delta E_{n}$ as a function of $n$. Note the different scales of the 
figures and the inset.}
\label{Fig6}
\end{figure}

We remind the reader that the classical SSO shows a crossover of its period as a function of the energy:
below the transition point $T(E)$ scales approximately linearly and decreases with increasing energy $E$,
whereas above this point it behaves nonlinearly and becomes increasing. Figure \ref{Fig6}(a) shows now the energy
eigenvalue spectrum as a function of the degree of excitation. Inspecting this spectrum one first
observes that there is an approximately linear dependence of the eigenvalues on the degree of excitation,
i.e. the spectrum is approximately equidistant.
This, on its own, is a suprising fact, since the potential of the SSO is very different from that of
the harmonic oscillator (see discussion in section \ref{sec:setup}) which possesses an equidistant spectrum of eigenvalues,
and since the eigenstates of the SSO and harmonic oscillator differ very much (see sections \ref{sec:gs} and \ref{sec:esa}). 
We remark that this approximate equidistance holds for the complete spectrum covering
low energies below the transition point as well as energies above the transition point up to
the last bound state, see Figure \ref{Fig6}(a) (note that the transition point is at
${\cal{V}}(q_{v}) \approx 3078$ which corresponds to the 20th excited state).
Let us next have a closer look at the spectrum of the SSO thereby revealing some intriguing features of it.

Figure \ref{Fig6}(b) presents the spacing of neighboring energy levels $\Delta E_{n+1} = E_{n+1} - E_{n}$
as a function of the
degree of excitation. We observe that the spacing consists of two branches. The first
branch of spacings (for low degrees of excitation) weakly linearly increases up to approximately
the 20th excitation. It subsequently starts to bend over and form the second branch which
decreases nonlinearly. This means that the crossover point encountered in the classical dynamics of
the SSO leaves its fingerprints also in the spacing of the energy levels of the quantum SSO.
The second branch for higher excitations possesses also a weakly pronounced 'irregular'
character. The decrease accelerates with increasing degree of 
excitation and, close to the maximum of the potential of the SSO, i.e. for the last
bound state, there is a strong sudden drop of the spacing: A diffuse 'threshold state' emerges.

The inset of Figure \ref{Fig6}(b) shows the scaled spacing $\Delta E_{n+1} / \Delta E_{n}$,
i.e. the ratio of energetically neighboring spacings,
as a function of the degree of excitation. This figure leads to an important observation.
For the first above-mentioned branch of excitations the scaled spacing is approximately 
constant and takes on the value $1.01$, i.e. the spectrum scales with a constant factor which deviates
weakly from the value one. When the crossover to the second branch happens, again close to the
20th excited state, the scaled spacing bends over and decreases strongly. Eventually,
for $n > 28$ pronounced fluctuations and oscillations occur.

\section{Excited eigenstate analysis} \label{sec:esa}

\begin{figure}
\parbox{10cm}{\hspace*{-2.6cm} \includegraphics[width=18cm,height=10.5cm]{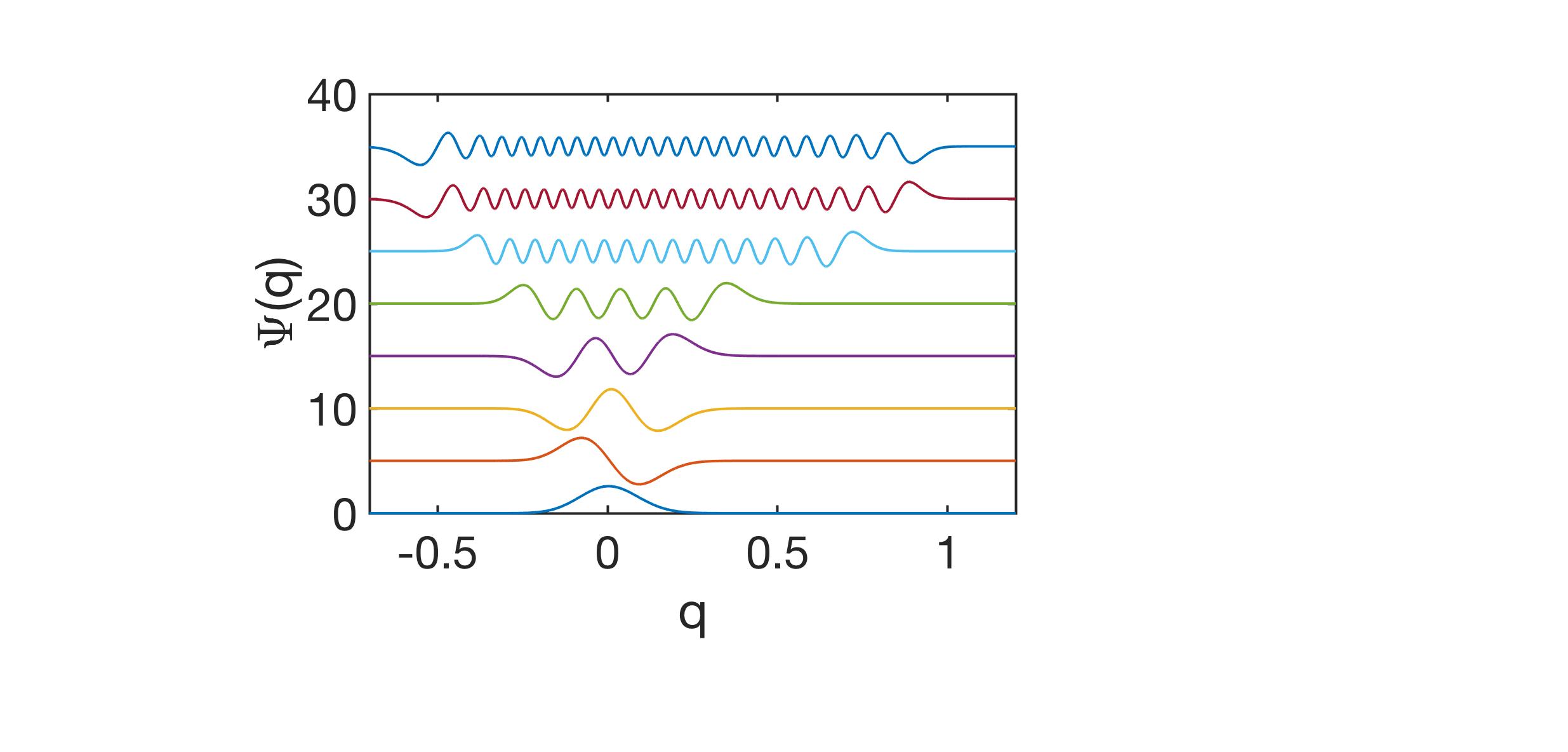}}
\vspace*{-1.4cm}
\caption{The ground state together with the 1st, 2nd, 3rd, 8th, 28th, 43rd and 44th excited eigenstates
for the amplitude $\alpha = 10^4$.}
\label{Fig7}
\end{figure}

We will now proceed and analyze the excited eigenstates of the SSO. Figure \ref{Fig7} shows the spatial
profile of the ground and selected excited eigenstates up to very high excitations close to the energetical
continuum for the amplitude $\alpha = 10^4$. The regular oscillatory and nodal structure is clearly visible.
Obviously, and as discussed above, there is no parity symmetry due to the missing reflection symmetry of
the SEP.

\begin{figure}
\parbox{10cm}{\hspace*{-0.6cm} \includegraphics[width=15cm,height=8.5cm]{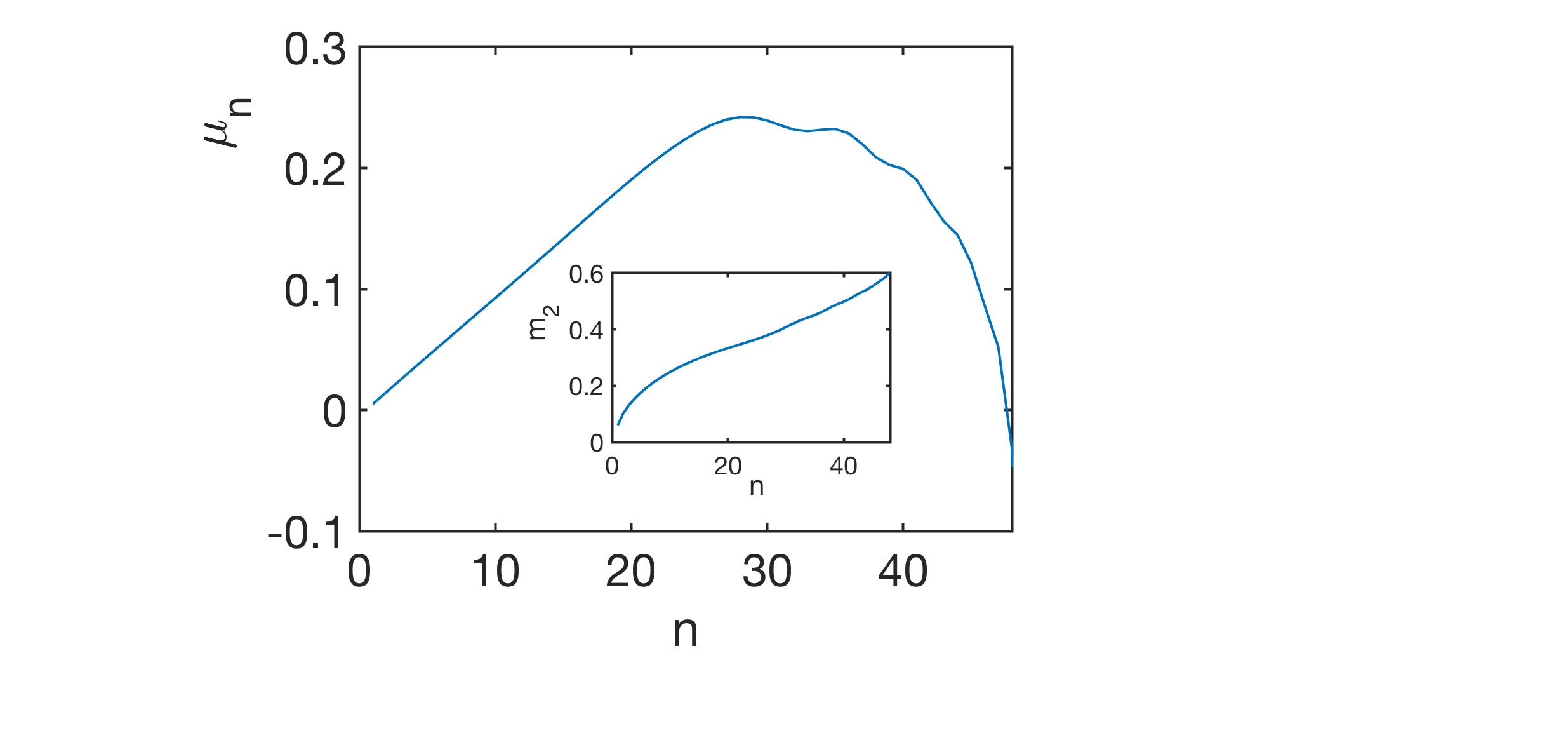}}
\vspace*{-1.4cm}
\caption{The mean $\mu_n=<\Psi_n|q|\Psi_n>$ with increasing degree of excitation for 48 bound states of the 
SSO. $\alpha = 10^4$. Inset: the variance $m_2$ with varying degree of excitation $n$. }
\label{Fig8}
\end{figure}

Figure \ref{Fig8} shows the mean $\mu_n=<\Psi_n|q|\Psi_n>$ with increasing degree of excitation $n$
for the 48 bound states of the SEP. 
We observe a linear increase of $\mu_n$ to positive values for low energy excitations i.e. within the first branch
mentioned above which corresponds to energies below the transition point.
For the second branch of excitations above the transition point a highly nonlinear behaviour of $\mu_n$
is encountered with an overall decrease finally turning to negative values for the highest excitations.
On top of the nonlinear decrease within the second branch we observe small amplitude oscillations.
This means that the decentering of the eigenstates for low excitations to positive values of $\mu_n$
and the right half of the SSO well (i.e. the part for $q>0$),
which is due to the inherent asymmetry of the SSO potential, turns
for higher excitations, due to the approaching of the maximum (or saddle point in phase space) into
a nonlinear decrease of $\mu_n$, i.e. a decrease of the degree of decentering, to finally an almost
zero decentering for the highest excited states. We note that even for the highest excited states
the majority of the nodes of the corresponding eigenstates are located on the right half of the well of the
SEP. The negative value of $\mu_n$ emerges then due to the tail development of the eigenstates in
the left outer part of the potential close to and beyond the maximum position (see Figs.\ref{Fig7} and 
\ref{Fig8}).

The inset of Figure \ref{Fig8} shows the variance $m_2$ for varying degree of excitation of the corresponding
eigenstates of the SSO. For low excitations a sublinear increase is observed (first branch, see above)
which turns over, following the transition point, to an approximately linear increase with some
small irregular appearing fluctuations on top. We therefore conclude with the statement that the spectrum, the
ground state and the excited states show a characteristic crossover in their behaviour when passing
the transition point of the SEP.

\section{Conclusions and outlook} \label{sec:conclusions}

We have explored and analyzed the spectral and eigenstate properties of the self-interacting
superexponential oscillator. While being of simple appearance the superexponential potential SEP
of this oscillator bears a remarkable variability. Of particular interest is the potential
well which has been investigated recently \cite{Schmelcher2} for the classical SSO. This potential
well is highly nonlinear and asymmetric and exhibits a transition point for which all derivatives
become singular. As a consequence the period of the classical oscillator as a function of its energy
undergoes a crossover from an approximately linearly decreasing to a nonlinearly increasing behaviour.

One of the central goals of this work is to find out whether the above crossover leaves its fingerprints in the
spectrum and eigenstates of the corresponding quantum oscillator. To this end we have first analyzed
the ground state with varying amplitude $\alpha$ of the SEP. While moving from low energies at the bottom
of the well towards the maximum energy at the continuum threshold the ground state undergoes a 
remarkable metamorphosis of decentering, asymetrical squeezing and the development of a tail.
Passing the energy of the transition point the ground state exhibits a characteristic shape change
which we analyzed in terms of the corresponding central moments. For low energies and large
values of the amplitude $\alpha$ the mean $\mu$ indicates a positive decentering which increases
with increasing energy (decreasing $\alpha$). At the transition point a maximum of $\mu$ occurs and
for higher energies a strong decrease leading to negative $\mu$ values is observed. The higher order
central moments show a characteristic transition from exponentially decaying moments to increasing
moments of higher order. 

The spectral analysis of the SSO showed the surprising fact that the energy eigenvalues in 
lowest order approximation (on the level of a percent accuracy) are equidistant all over the spectrum
from the bottom of the well to its maximum. An equidistant spectrum, known from the harmonic oscillator, 
is a very unexpected behaviour for the SSO, since the SEP is extremely nonlinear even when compared
to nonperturbative anharmonic oscillators. A closer inspection by analyzing the scaled
spacing of the energies, however, revealed that there is a scaling of the spectrum for energies 
below the transition point energy which turns into a nonlinear irregularly fluctuating decrease for
energies above the transition point. Therefore, the two-fold classical dynamics in the segmented SEP
leaves some clear and pronounced fingerprints both in the energy spectrum as well as the profile of the
eigenstates. We have shown this also for the excited eigenstates up to the last bound so-called
'threshold state'.

The prospects of further investigations are as follows. The facts of the approximate equidistance
of the spectrum and the very different eigenfunctions as compared to standard power law oscillators 
(see appendix A) suggests that our SSO could show approximate revival and recurrence features in the
corresponding quantum dynamics. To explore the latter is therefore an interesting and immediate
perspective resulting from the present work. Coherent, non-dispersive and squeezed states of light,
being a harmonic oscillator in amplitude space, play a major role and are of immediate importance
in modern physics. One might therefore pose the question whether the SSO exists in the form of a corresponding 
realization for light-matter systems (see remarks below).

The present work provides some key ingredients for following the pathway from a single to many
coupled oscillators. This pathway has proven to be of outstanding importance in understanding
the nonlinear and localized excitations as well as traveling wave solutions such as solitons.
It is therefore an intriguing perspective to explore the generalization of the SSO for e.g. chains of nonlinear 
oscillators.

Finally, we would like to remark on a possible experimental realization of the SSO. In ultracold
atomic physics and Bose-Einstein condensates there exist unprecedented possibilities to control
the atomic motion and the atomic interactions. Traps of practically arbitrary geometry can be
designed and prepared experimentally \cite{Pethick,Grimm}. One particularly flexible strategy are
the so-called painted arbitrary and dynamic potentials based on a rapidly moving laser beam
that 'paints' a time-averaged optical dipole potential \cite{Henderson}. The long coherence
times of ultracold gases mostly due to the suppression of three-body recombination related
to their dilute character i.e. low densities renders them a good platform for the realization
of the SSO and its quantum features. Due to the fact that single atom loading and detection is possible
\cite{Ott}, or, alternatively, switching off the interactions in the many-body case via magnetic
Feshbach resonances \cite{Chin} the corresponding single particle physics could be probed.

\section{Acknowledgments} \label{sec:acknowledgments}

The author acknowledges helpful discussions with F.K. Diakonos and a careful reading of the
manuscript by C.V. Morfonios.

\appendix

\section{Some facts on power law oscillators} \label{sec:appendix}

In order to contrast the behaviour of the SSO as explored and analyzed in this 
work with the standard power law oscillator, we provide here a brief summary of
some of the main features of the power law oscillators \cite{Bender,Znojil,Quigg,Liverts}.
We consider the power law oscillators with continuously varying exponents

\begin{equation}
{\cal{H}} = \frac{p^2}{2m} + \frac{1}{2} \alpha |q|^{\beta}, \alpha,\beta>0
\end{equation}

This covers all fractional powers and includes in particular the linear, quadratic and higher order anharmonic
potentials up to $\beta \rightarrow \infty$ where we reach the box-like confinement.
Let us address some spectral features of this class of oscillator potentials.
A corresponding semiclassical analysis is based on the quantization condition

\begin{equation}
\int_{q_1}^{q_2} p(q) dq = \left( n + \frac{1}{2} \right) \pi
\end{equation}

where $q_1,q_2$ are the coordinates of the turning points of the one-dimensional oscillatory motion.
The $\frac{\pi}{2}$ phase is appropriate for a smooth behaviour of the potential around
the turning points, and has to be replaced by a phase of $\pi$ for the hard wall boundary
conditions.  This yields for the eigenenergies \cite{Quigg} (using $\alpha = m = 1$)

\begin{figure}
\parbox{10cm}{\includegraphics[width=14cm,height=7.5cm]{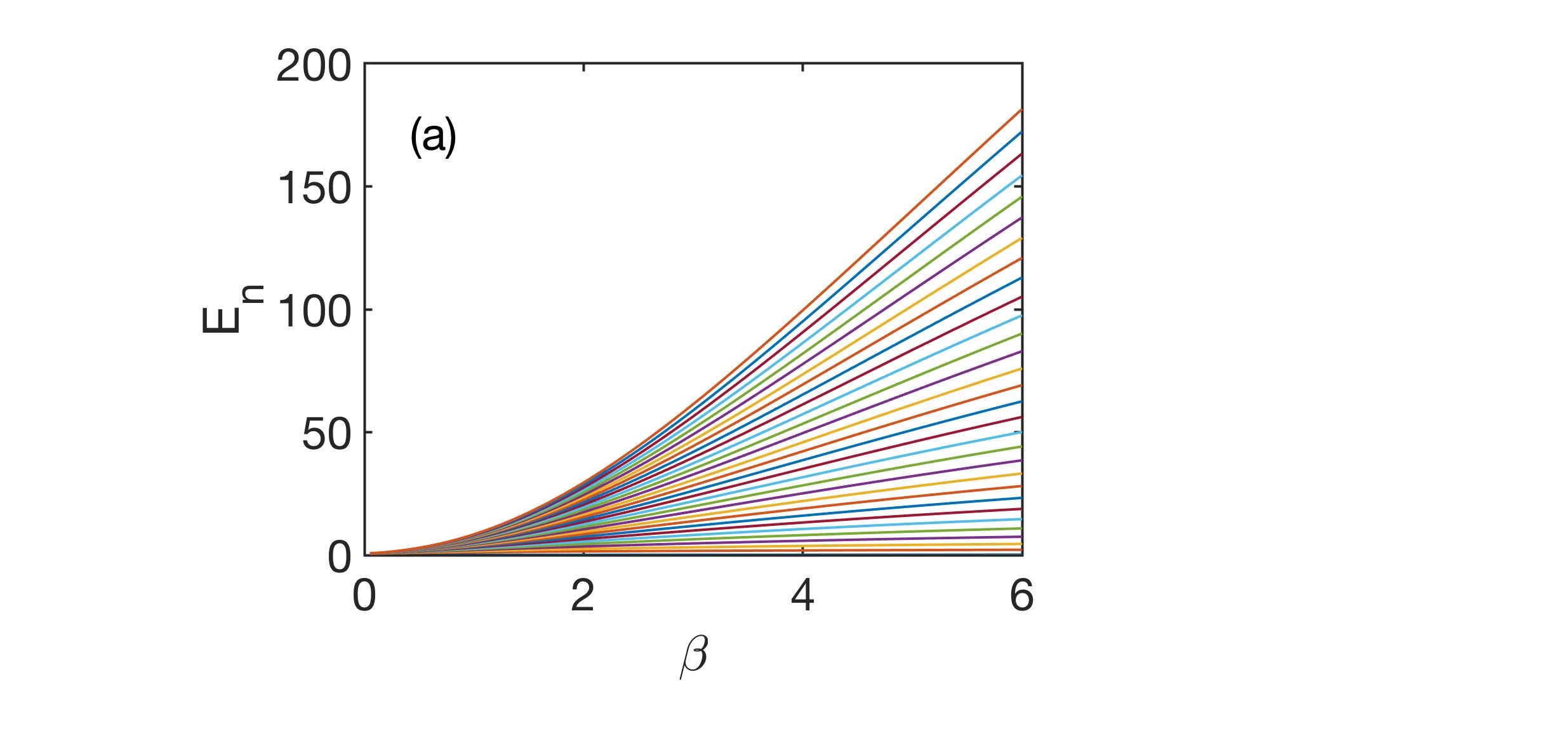}}
\parbox{10cm}{\hspace*{1cm} \includegraphics[width=13.4cm,height=7.5cm]{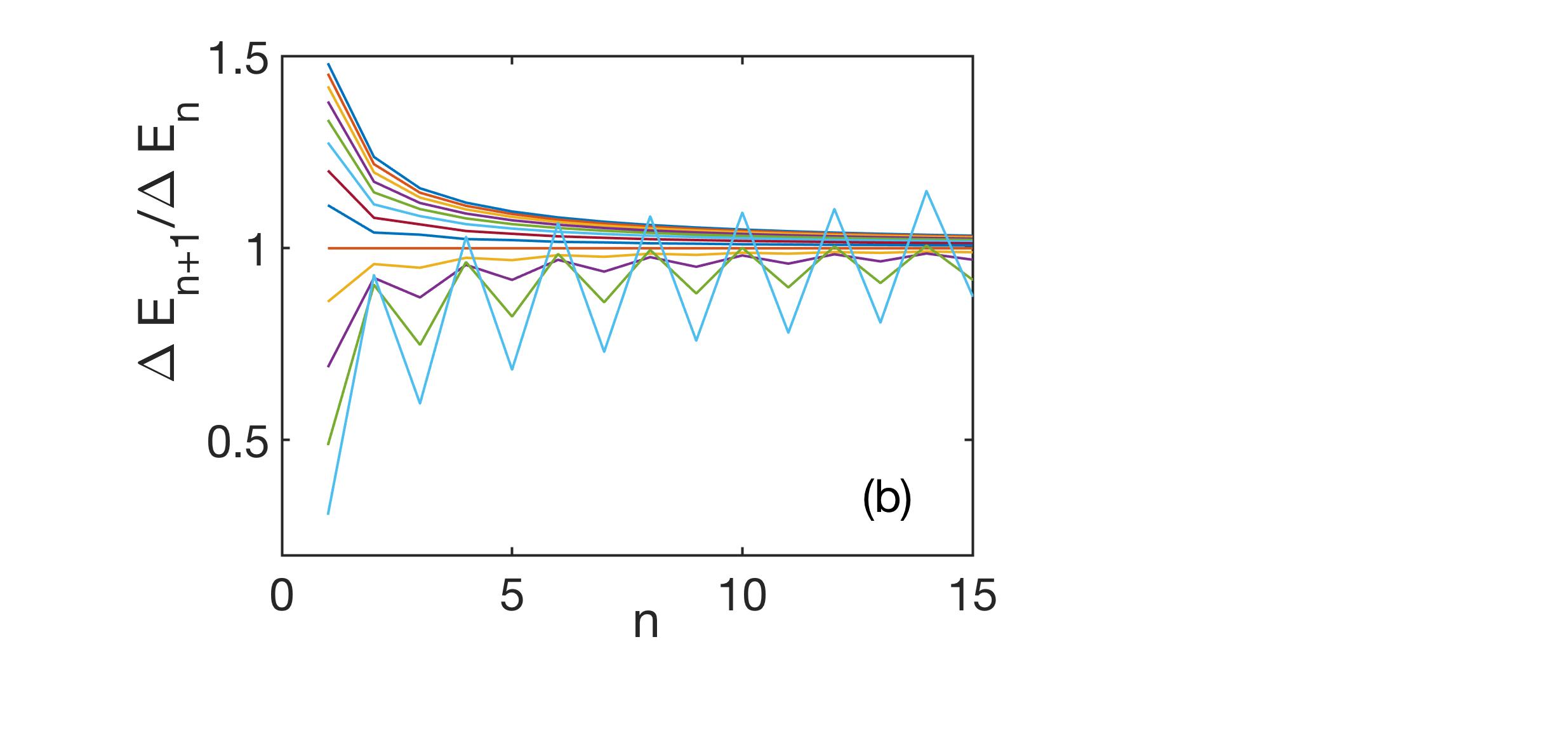}}
\caption{(a) The energetically lowest 30 eigenvalues $E_n (\beta)$ of the spectrum of the power law oscillators
with varying exponent $\beta = 0.05-6$. (b) The scaled spacing $\Delta E_{n+1} / \Delta E_{n}$
as a function of the degree of excitation $n$ for the values $\beta$ ranging from $6$ to $0.5$ in steps
of $0.5$ and for the value $0.1$. Order from top to bottom corresponding to a decreasing value of $\beta$.}
\label{Fig9}
\end{figure}

\begin{equation}
E_n = \frac{1}{2} \left( \left(n+\frac{1}{2} \right) \sqrt{\pi} \frac{\Gamma \left(\frac{3}{2} +\frac{1}{\beta} \right)}
{\Gamma \left(1 +\frac{1}{\beta} \right)} \right)^{\frac{2 \beta}{2+\beta}}
\end{equation}

The lowest thirty energy eigenvalues of a numerically exact eighth order finite difference calculation
are shown in Figure \ref{Fig9}(a). The energy eigenvalues are shown as a function of the
exponent $\beta$ ranging from $0.1$ to $6$. The spectrum becomes increasingly dilute with
increasing exponent $\beta$ and, except for the ground state, the energies depend 
monotonically on $\beta$. The spacing $\Delta E_{n+1} = E_{n+1} - E_{n}$
(not shown here) of the energy levels is monotonically increasing for $\beta > 2$ whereas
it is monotonically decreasing for $\beta < 2$. For a fixed degree of excitation the spacing
increases with increasing value of $\beta$. The scaled energy spacing $\Delta E_{n+1} / \Delta E_{n}$
is shown in Figure \ref{Fig9}(b) where the different curves correspond to different selected
values of $\beta$. Asymptotically, i.e. for $n \rightarrow \infty$, all scaled spacings
approach the value one. However, the spectral behaviour shows major deviations from this
asymptotic limit in particular for the part considered here. The scaled spacing
is always larger than one for $\beta > 2$ and monotonically decreasing for increasing degree
of excitation. For $\beta < 2$ the scaled spacing is predominantly smaller than one but
it shows with a decreasing value of $\beta$ a more and more pronounced nonmonotonous i.e.
alternating behaviour with increasing degree of excitation.

All of the above clearly demonstrates that the spectral properties of the power law potentials
for arbitrary but constant power are very different from the spectral properties of the SSO.


\begin{thebibliography}{99}
\bibitem{Scully} M.O. Scully and M.S. Zubairy, Quantum Optics, Cambridge University Press (1997).
\bibitem{Arenas} A. Arenas {\it{et al}}, Phys.Rep. 469, 93 (2008).
\bibitem{Ashcroft} N.W. Ashcroft and N. Mermin, Solid State Physics, Harcourt College Publishers (1976).
\bibitem{Wilson} E. Bright Wilson Jr., J.C. Decius and P.C. Cross, Molecular Vibrations, Dover Books on Chemistry, 1980.
\bibitem{Flach1} S. Flach and C.R.Willis, Phys.Rep. 295, 181 (1998).
\bibitem{Flach2} S.Flach and A.V.Gorbach, Phys.Rep. 467, 1116 (2008).
\bibitem{Kevrekidis} P.G. Kevrekidis, D.J. Frantzeskakis and R. Carretero-Gonzalez, The Defocusing Nonlinear Schr\"odinger
Equation, Society for Industrial and Applied Mathematics, United States (2016).
\bibitem{Kartashov} Y.V. Kartashov, B.A. Malomed, Lluis Torner, Rev.Mod.Phys. 83, 247 (2011).
\bibitem{Cao} L.S. Cao {\it{et al}}, Phys.Rev.Lett. 112, 075505 (2014).
\bibitem{FPU} The Fermi-Pasta-Ulam Problem, edited by G. Gallavotti, Lecture Notes in Physics 728
(Springer-Verlag, Berlin, 2008).
\bibitem{Casati} G. Casati, B.V. Chirikov, D.L. Shepelyansky, I. Guarneri, Phys.Rep. 154, 77 (1987).
\bibitem{Izrailev} F.M. Izrailev, Phys.Rep. 196, 299 (1990).
\bibitem{Stoeckmann} Quantum Chaos: An Introduction, H.J. St\"ockmann, Cambridge University Press (1999).
\bibitem{Fossen} T.I. Fossen and H. Nijmeijer, Parametric Resonance in Dynamical Systems, Springer (2011).
\bibitem{Arbell} H. Arbell and J. Fineberg, Phys.Rev.E 65, 036224 (2002).
\bibitem{Szwaj} C. Szwaj {\it{et al}}, Phys.Rev.Lett. 80, 3968 (1998).
\bibitem{Schmelcher1} P. Schmelcher, Phys.Rev.E 98, 022222 (2018).
\bibitem{Schmelcher2} P. Schmelcher, J.Phys.A 53, 075701 (2020).
\bibitem{Schmelcher3} P. Schmelcher, arXiv:2001.07705, subm. Phys.Rev.E.
\bibitem{Bender} C.M. Bender and T.T. Wu, Phys. Rev. 184, 1231 (1969).
\bibitem{Znojil} M. Znojil, J.Phys.A 15, 2111 (1982).
\bibitem{Quigg} C. Quigg and J.L. Rosner, Phys.Rep. 56, 167 (1979).
\bibitem{Liverts} E.Z. Liverts, V.B. Mandelzweig and F. Tabakin, J.Math.Phys. 47, 062109 (2006).
\bibitem{Groenenboom} G.C. Groenenboom and H.M. Buck, J.Chem.Phys. 92, 4374 (1990).
\bibitem{Pethick} C.J. Pethick and H. Smith, Bose-Einstein condensation in dilute gases, Cambridge University Press (2008).
\bibitem{Grimm} R. Grimm, M. Weidem\"uller and Y.B. Ovchinnikov, Adv.At.Mol.Opt.Phys. 42, 95 (2000).
\bibitem{Henderson} K. Henderson {\it{et al}}, New J.Phys. 11, 043030 (2009).
\bibitem{Ott} H. Ott, Rep.Prog.Phys. 79, 054401 (2016).
\bibitem{Chin} C. Chin {\it{et al}}, Rev.Mod.Phys. 82, 1225 (2010).
\end{thebibliography}
\end{document}